\documentclass[aps,prb,preprint,groupedaddress]{revtex4-1}
 \usepackage{graphicx}

\begin{document}


\title{Defect Induced Resonances and Magnetic Patterns in Graphene}
\author{Yi Chen Chang}
\email{changyic@usc.edu}
\author{Stephan Haas}
\affiliation{Department of Physics and Astronomy, University of Southern California, Los Angeles, CA 90089-0484 }
\date{\today}

\begin{abstract}
We investigate the effects of point and line defects in monolayer graphene
within the framework of the Hubbard model, using a
self-consistent mean field theory. These defects are found to induce
characteristic patterns into the electronic density of states and cause
non-uniform distributions of magnetic moments in the vicinity of the
impurity sites. Specifically, defect induced resonances in the
local density of states are observed at energies close to the Dirac points.
The magnitudes of the frequencies of these resonance states are shown to
decrease with the strength of the scattering potential, whereas their
amplitudes decay algebraically with increasing distance from the defect.
For the case of  defect clusters, we observe that 
with increasing defect cluster size the local magnetic moments in the 
vicinity of the cluster center are strongly enhanced.
Furthermore, non-trivial impurity induced magnetic patterns are observed in
the presence of line defects: zigzag line defects are found to introduce
stronger-amplitude magnetic patterns than armchair
line defects. When the scattering strength of these topological defects is
increased, the induced patterns of magnetic moments become more strongly
localized.
\end{abstract}


\maketitle

\section{introduction}
In situ formation of atomic size defects has recently been observed in
graphene layers, using transmission electron microscopy.\cite{H04}
Specifically, it was demonstrated that certain topological defects can be
induced by irradiation with electrons beams, thus raising the possibility
that more complex impurity structures, such as specifically tailored line
defects, can in principle be achieved using similar experimental techniques.
The effects of such designer impurity structures on the nanoscale are
interesting, as they can have profound effects on the electronic properties
of the material. Similar to anisotropic superconductors, graphene is known
to have a reduced electronic phase space close to its Fermi surface,
resulting in a linear low-energy density of states.\cite{AVB06} For the case
of unconventional superconductors, the introduction of such defects has been
shown to have profound consequences, such as formation of low-energy bound
states (or quasi-bound states) and localization of quasiparticles.\cite{SH00} In this paper we show
that these features have an analogue in metallic graphene, in particular
when magnetic fluctuations are taken into account.

The effects of impurities in graphene are of particular interest because
their presence has been shown to strongly reduce the otherwise remarkably
high electronic mobility in this compound\cite{FS07,SA07} and to change its
electronic band structure\cite{RNC06,HS10}. Furthermore, impurity induced
local puddles of charge carriers have been proposed to be responsible for
the observed minimum conductivity.\cite{SA07} Very recently, impurity
induced bound states have been experimentally observed using scanning
tunneling microscopy. \cite{MMU10} Specifically, it was shown that the
tunneling current amplitude of these single impurity bound states decays
inversely with the square of the distance from the defect. Because of this
algebraic dependence, they are in fact quasi-bound states. It was also
suggested that such defects induce local magnetic moments\cite{HS10,ADH10},
which in turn can cause global ferromagnetic instabilities with a transition
temperature that scales as the square root of the impurity concentration. 
\cite{MMU10}

A number of properties of graphene sheets with point impurities have already
been established \cite{VMP06,MMU10,OVY07,NMRP06}. In particular, in a recent
experiment extended one-dimensional defects were realized \cite{JL10},
demonstrating that the creation of designer defects in graphene sheets is
becoming realistic. In this paper, we investigate the effects of such
topological defects on the local densities of state and on the magnetization
patterns. Specifically, we study the impurity-doped Hubbard model\cite%
{NMRP04} on a graphene sheet geometry, where we consider the cases of single
impurities and one-dimensional impurity clusters.

This manuscript is organized as follows. In the following section, we
discuss the model, the approximations used, and the quantities we
investigate. In the subsequent section, we show results for the induced
density of states and for the magnetization in the impurity doped Hubbard
model. We conclude with a section summarizing our results and discussing
possible experimental implications. 

\section{Model}

We consider the Hubbard model in the presence of non-magnetic impurities,
given by the Hamiltonian 
\begin{equation}
H=t\sum_{<i,j>,\sigma }^{N}(c{_{i\sigma }^{\dagger }c_{j\sigma }+h.c.)}%
+U\sum_{i}^{N}n_{i\uparrow }n_{i\downarrow }+U_{d}\sum_{i,\sigma }^{N_{defect}}(c{%
_{i\sigma }^{\dagger }c_{i\sigma }+h.c.)},
\end{equation}%
where the sums extend over the two-dimensional  honeycomb  lattice
t=-2.7eV is the orbital hopping integral, and 
$c_{i\sigma },c_{i\sigma }^{\dagger }$ are
electron creation and annihilation operators respectively. The second term
denotes the on-site Coulomb repulsion, where $n_{i\sigma }$ is the number
operator. The third term describes the scattering of the electrons by local
defects. When a strong impurity, such as a vacancy, is created, the
scattering strength $U_{d}$ goes towards infinity. In the following, we
treat this model within a mean field approximation, i.e. the Coulomb
repulsion term is approximated as 
\begin{equation}
H_{mf}=U\sum_{i}^{N}\left( \left\langle n_{i\sigma }\right\rangle n_{i-\sigma 
}-\left\langle n_{i\sigma }\right\rangle \left\langle n_{i-\sigma
}\right\rangle \right).
\end{equation}%
We choose U=1.2t, leading to a semi-metallic phase with a conical
dispersion, as observed in graphene. This parameter choice is below the
critical value $(U/t)_c=2.2$\cite{FS07,SA07} at which there is a quantum
phase transition to an antiferromagnetic insulating phase. The mean charge
density $%
\left\langle n_{i\sigma }\right\rangle $ is computed self-consistently from 
\begin{equation}
\left\langle n_{i\sigma }\right\rangle =\int dEg_{i\sigma }(E)f(E-E_{f}),
\end{equation}%
where $g_{i\sigma }(E)=\sum_{j}\Psi _{i}^{\ast }(E_{j})\Psi
_{i}(E_{j})\delta (E-E_{j})$ is the local electronic density of states, and $%
f(E-E_{f})$ is the Fermi function. This self-consistent solution provides the
local densities of states and the spin densities $M_{i}=(\left\langle
n_{i\sigma }\right\rangle -\left\langle n_{i-\sigma }\right\rangle )/2$ on
each atom. The calculations discussed in the following section are performed
in real space on finite 960-site honeycomb lattices.

\section{Results and discussion}
\subsection{Single vacancy}

Let us start by considering the effects of a single impurity in graphene.
Fig. 1(a) shows the calculated zero-energy local density of
states for a graphene sheet in the vicinity of a vacancy. A localized
state is observed to form in the vicinity of the defect site,
as has been reported previously\cite{VMP08,VMP06}, with a
characteristic triangular spatial pattern that is commensurate with the
lattice symmetry. For the parameters chosen here, i.e. $U_{d}/t=1000$
(corresponding to a vacancy), the
energy of the induced bound state is at -0.1eV. In Fig. 1(b),
we compare the local densities of states at a site next to the vacancy and
at another site far away from it. The bound state is clearly absent in the
latter case, which instead shows the well known Dirac cone shape. Note that
the linear dispersion is slightly smeared out by the finite broadening ($%
\gamma $/t=0.083) obtained as the delta functions in ${g}_{i\sigma }(E)$ are
replaced by Lorentzians.

 \begin{figure}
\scalebox{0.6}{\includegraphics{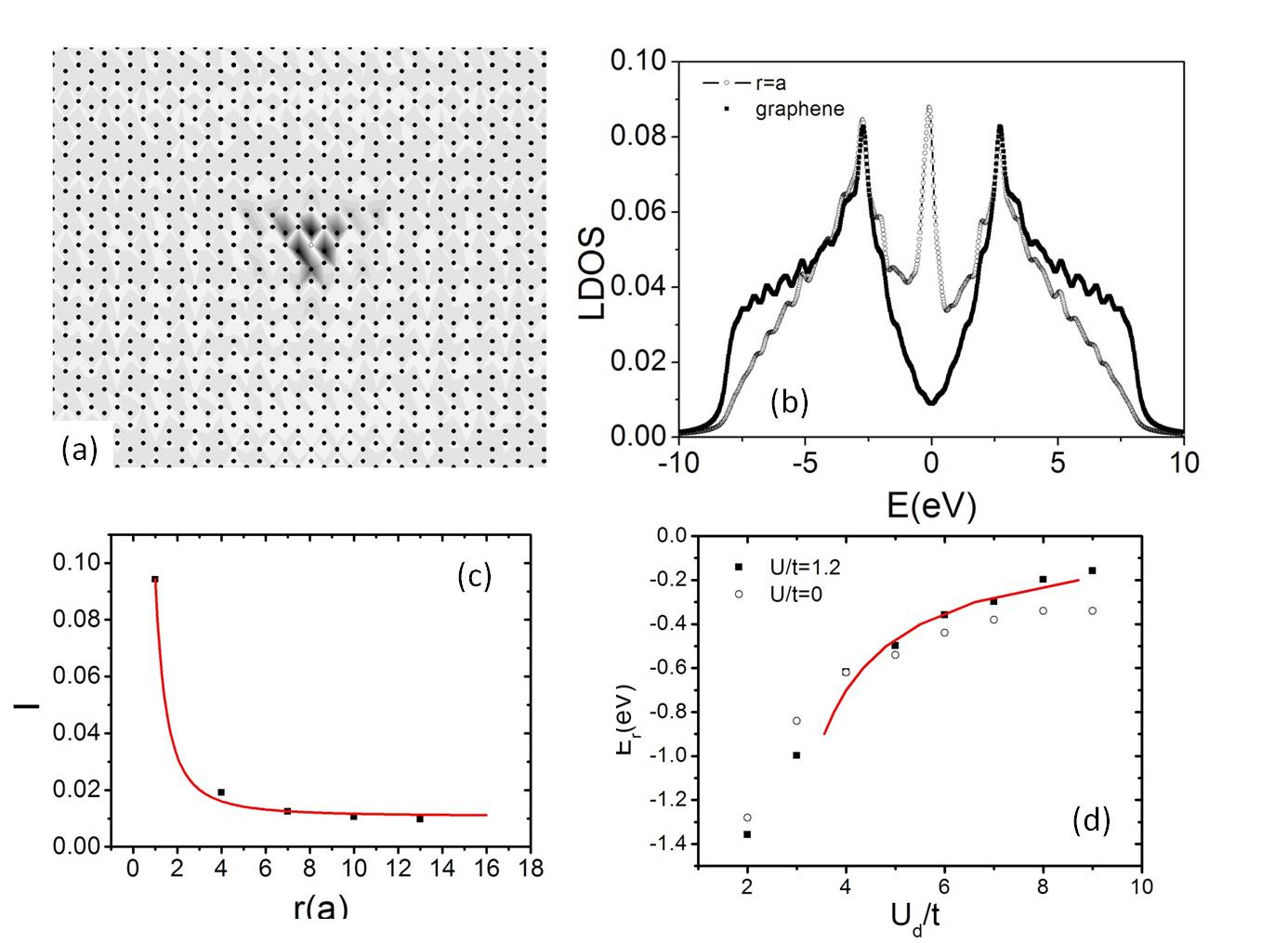}}
 \caption{(a) Zero-energy local density
of states in a graphene sheet with a single vacancy. (b) Local density of
states at a lattice site next to the vacancy (open circles) and at a site
far away from it (black solid line). (c) Spatial dependence of the intensity
of the low-energy peak in the local density of states, corresponding to an
impurity induced quasi-bound state. The solid red curve is a fit to a $r^{-2}$
decay. (d) Resonance energy at the impurity site as a function of the
scattering strength $U_{d}$. The red solid curve is a fit to the asymptotic
regime discussed in the text, yielding a bandwidth W =
5.6eV}
 \end{figure}

Next we examine the spatial decay of the amplitude of the defect induced
bound state. As observed in Fig. 1(c), the magnitude of the
impurity peak can be fitted well by a power law proportional to the squared
inverse of the distance from the vacancy, which originates from the r$^{-1}$
decay of the bound state wave function\cite{VMP06}, and which is hence
actually a quasi-bound state with power-law decay. This is the same
algebraic decay which has recently been reported by scanning tunneling
experiments\cite{MMU10}. Furthermore, similar power-law decay has been
observed for quasi-bound states around non-magnetic impurities in anisotropic
superconductors along certain directions\cite{AVB95}. In Fig. 1(d), we study the dependence of the resonance energy on the
magnitude of the impurity scattering strength. The observed dependence is in
agreement with the resonant scattering behavior reported by Skrypnyk et. al.%
\cite{YVS06} and Wehling et. al \cite{TOW07}, i.e. with the resonance energy
($E_{r}$) satisfying 
\begin{equation}
U_{d}=\frac{W^{2}}{E_{r}ln\left\vert \frac{E_{r}^{2}}{W^{2}-E_{r}^{2}}%
\right\vert },
\end{equation}%
where W is the bandwidth. When a vacancy is created, $U_{d}/t$ is infinite,
and the resonance peak is close to the Dirac point. A fit to the
asymptotic behavior $E_{r}=1-exp(\frac{W^{2}}{2U_{d}})$ in the strong
coupling regime indicates that for the present case the bandwidth is 5.6eV.%
\cite{note} When the on-site Coulomb repulsion U/t is increased, we find
that the resonance peak continues to follow the same dependence as in Fig.%
1(d). The fit shown here is only for the regime of scattering
strengths larger than $U_{d}/t=4$.

\subsection{Line vacancies}

Next, let us examine the effects of more extended defects in graphene on the
electronic density of states. In Fig. 2 we study the local
density of states in the vicinity of a zigzag line defect, consisting of
13 contiguous vacancies in a graphene sheet. The Hamiltonian parameters are
chosen to be the same as for the point defect discussed above.
Similar to the case of a single vacancy, we observe that the local density of states
shows a pronounced low-energy peak close to the impurity, indicating
localization of the charge carriers. As expected, the amplitude of this
induced resonance peak in the local density of states is enhanced with
respect to the case of a single impurity. For example, at the central site
next to the line defect for the given parameters it is amplified
approximately by a factor of 2.2, a direct consequence of constructive
interference of the joint point defects. Furthermore, we find that the
spatial decay in the local density states at the Dirac point is the same as
for the point defect when moving away from the line defect in a
perpendicular direction. 
As shown in the upper inset of Fig. 2(a), it falls off with an inverse square power law along the path
indicated in the lower inset. 

\begin{figure}
\scalebox{0.6}{\includegraphics{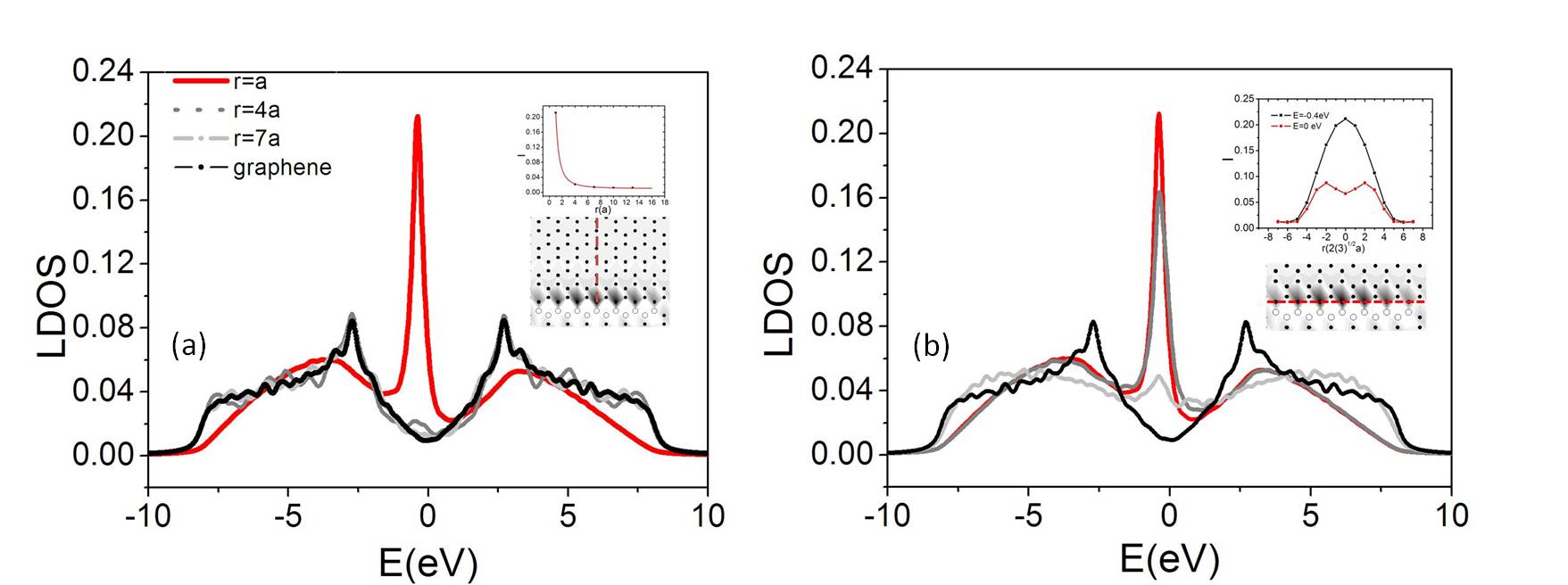}}
 \caption{(a) Local density of states at
various distances from a zigzag line defect in a graphene sheet. The upper
inset shows the intensity of the zero energy local density of states as a
function of the position along the direction perpendicular to the defect,
indicated by the red dash line in the lower inset. (b) Same as part (a), but
along the direction parallel to the defect, as shown in the lower inset.  }
 \end{figure}

In contrast, the zero energy local density of
states parallel to the line defect does not behave monotonically. In Fig. %
2(b), we plot the local density of states along a path parallel
to the line defect. It is evident that the magnitude of the defect induced
bound state varies by two orders of magnitude along this cut. Furthermore,
in contrast to the perpendicular direction, an M shaped dependence in ${g}%
_{i\downarrow }(E=0)$ is observed, indicating destructive interference close
to the center of the defect, and maxima at two non-trivial positions along
the parallel cut. At the resonance energy, on the other hand, ${g}%
_{i\downarrow }$(E$_{r}=-0.4eV$) shows Gaussian behavior, with the maximum
located at the center of the impurity cluster, indicating constructive
interference.

Analogous bound state phenomena have recently
been observed close to zigzag edges\cite{Wk99,MAHV05,KWM96}.

\subsection{Impurity clusters and random defects}

The shape and  size of impurity clusters is known to  profoundly affect  the
conductivity in the graphene sheet. In particular, it is believed that the carrier
concentration is dramatically modified  by the presence of
extended defects in the system\cite%
{AA09}. Here we examine the specific case of one-dimensional zigzag 
vacancy clusters with variable length. In Fig. 3, the effects of defect cluster size 
on the local and global density of states are shown, using the same parameter 
choices as in the previous sections. 

\begin{figure}
\scalebox{0.7}{\includegraphics{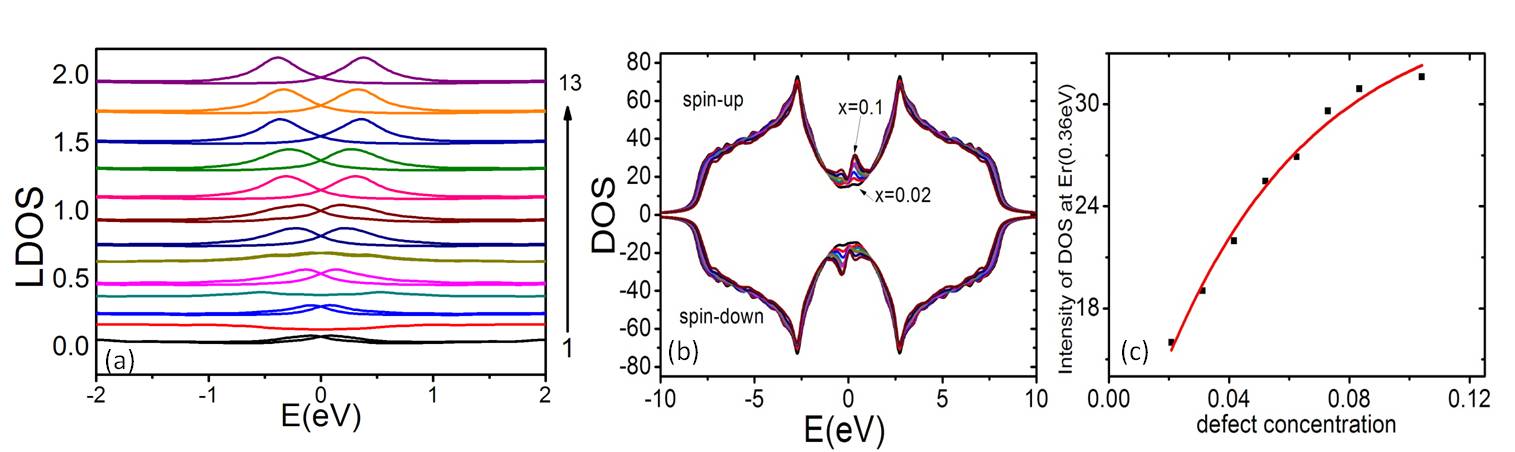}}
 \caption{(a) Low-energy local density of states (LDOS) in the vicinity of linear impurity clusters
 of varying size, ranging from a single vacancy (bottom) to 13
vacancies (up). The LDOS is measured at a lattice site next to the center of the
defect chain. The resonance peaks at positive energies correspond to spin-up
electrons, whereas the negative resonances correspond to spin-down electrons;
(b) global density of states for different concentrations of randomly placed
defects; (c) intensity of the resonance peak in (b), located at 0.36eV, as a
function of defect concentration
}
 \end{figure}

As seen in Fig. 3(a), when the number of vacancy sites (n$_{d}$) in the impurity cluster
is gradually increased from 1 to 13, a significant feature in the local density of states
is that the intensity of the resonance peaks is enhanced due to the formation of an
impurity band in the vicinity of the Dirac point. 
This is in agreement with a prediction of Pereira et. al.\cite%
{VMP06}. Second, the spin-up resonance is blue-shifted with increasing
size of the impurity cluster, whereas the spin-down resonance is red-shifted.
This is due to the presence of on-site Coulomb
interactions, leading to stronger spin polarization in the vicinity of
vacancies cluster, as we will discuss in the following section.
Notice also that the resonance peaks are not present in the smallest 
impurity clusters with n$_{d}$=2 and 4, reflecting the
absence of induced local magnetic moments. This striking effect, which is 
visible only in the smallest impurity clusters, 
is due to the cancellation of defect induced spin-polarized bound states, in
agreement with an observation recently reported by Kumazaki et. al.\cite{HK08}

Next, we focus on the case where vacancies are randomly distributed 
within in the graphene sheet. Fig. 3(b) shows the global density of states as a
function of defect concentration. In the low-energy region, the intensity of the
resonance peak (E$_{r}\sim$0.36 eV) is found to be significantly enhanced with 
increasing defect concentration. This indicates that this feature in the global 
density of states is not just a simple superposition of the local density of states 
surrounding the vacancies. The space between the vacancies is reduced
when the defect concentration increased, and the contributions from the localized
states are therefore reduced. This implies that a maximum conductivity will be
observed when a critical defect concentration is formed in the graphene sheet.
Notice that, another relatively small resonance peak located at
E$_{r}\sim$-0.36 eV, which originates from the uneven number of spin up and
spin down electrons in the system, resulting from the random distribution of
vacancies sites. In the high energy region, a  softening of the van Hove
singularity located at $\pm2.7eV$ and  development of Lifshitz tails at the
band edge are induced by increasing the defect concentration\cite{VMP08}.%

\subsection{Magnetic patterns}

Next, we examine the magnetic patterns induced by defects. Within the
self-consistent mean field calculation the numbers of spin-up and spin-down
electrons are fixed during the iteration process, whereas the total number
of electrons is kept equal to the number of carbon atoms. As the honeycomb
lattice of graphene is composed of two sublattices, containing atoms A and
B, the presence of a single vacancy defect implies that the numbers of A
atoms (N$_{A}$) and B atoms (N$_{B}$) are not equal. Therefore magnetic
moments are induced, consistent with Lieb's theorem\cite{Lieb}, i.e. the
total spin of the ground state is S = (N$_{A}$ -N$_{B}$)/2.

Let us first examine the magnetic pattern in a graphene sheet induced by a
single defect, shown in Fig.4(a). The total magnetic
moment in this case is 0.5/960, since only one A atom in the 960-site sheet
is missing. The magnetic moment is localized around the vacancy resembling
the LDOS shown in Fig. 1(a)\textbf{.} If the vacancy is
introduced into the A sublattice, the magnitude of the induced magnetic
moment in the at B sublattice is larger than in the A sublattice with a
maximum value $M_{B}=-0.058\mu _{B}$ and $M_{A}=0.0088\mu _{B}$. This
indicates that the effective magnetic interaction between spins in opposite 
sublattices is
antiferromagnetic, and the interaction between spins on the same sublattice
is ferromagnetic. \cite%
{MPL09,HK08,AMB10,Moh}.

 \begin{figure}
\scalebox{0.7}{\includegraphics{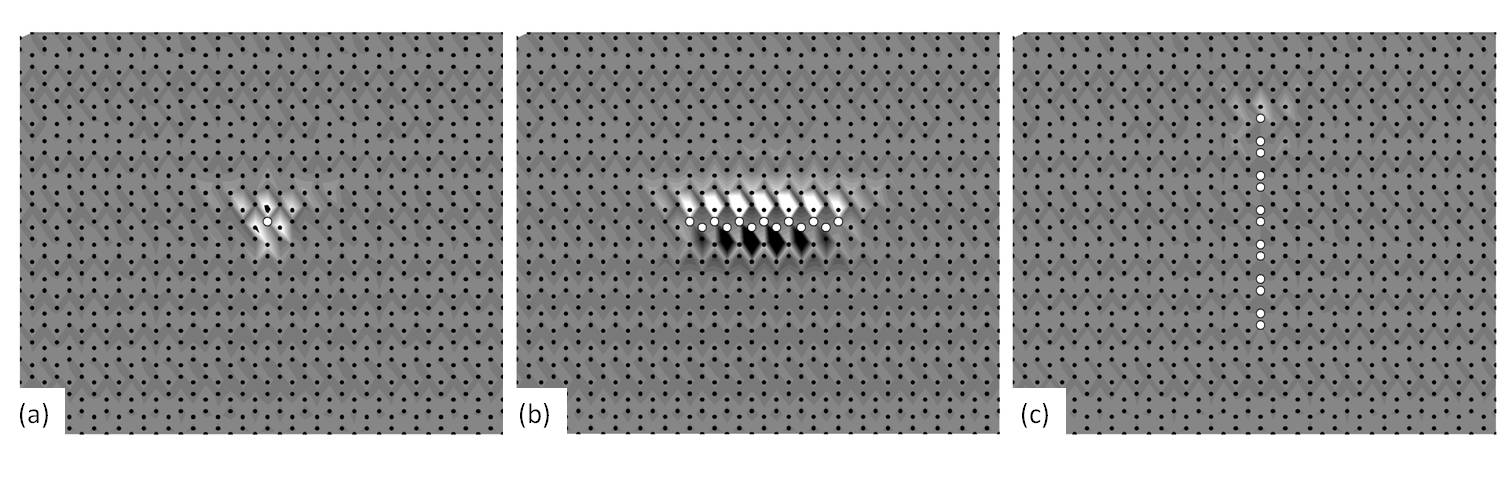}}
 \caption{Magnetic patterns in a graphene
sheet with by (a) a single defect, (b) a zigzag type line defect containing
13 atoms, and (c) an armchair type line defect, also containing 13 atoms.
The defects are placed in the center of the sheet, as indicated by the white
symbols. The color scale ranges from white (strong spin-down
magnetization) to black (strong spin-up magnetization). The scattering
strength is $U_{d}/t=1000$, corresponding to vacancies. }
 \end{figure}

When a zigzag type line defect is introduced, a pronounced localized
magnetic pattern is formed close to the defect, as observed in Fig.4(b). 
For example, when taking out 7 A atoms and 6 B atoms, the induced
magnetic moment magnitudes in the B sublattice are larger than the A
sublattice with maximum impurity induced magnetic moment $\ M_{B}=-0.151\mu _{B}$ and $%
M_{A}=0.146\mu _{B}$. Although the total magnetic moment remains 0.5/960,
the local magnetic moment is 2.6 times larger than for a single vacancy in
the spin-down case and 16.6 times larger for the spin-up case. This can be
understood by comparing the intensities of the LDOS resonance peaks in these
cases. The magnetic moment shows a spatial Gaussian shape, following the
resonance in the local density of states shown in Fig. 2(b). 

Fig.4(c) shows the impurity induced magnetic pattern of a graphene sheet in
the presence of an armchair line defect. Compared to the zigzag line defect
and the single defect, the induced local magnetic moment is much weaker.
This indicates the absence of impurity bound states in the vicinity of 
armchair line defects and armchair edges, as also reported in 
previous work\cite{Wk99}.  

\begin{figure}
\scalebox{0.7}{\includegraphics{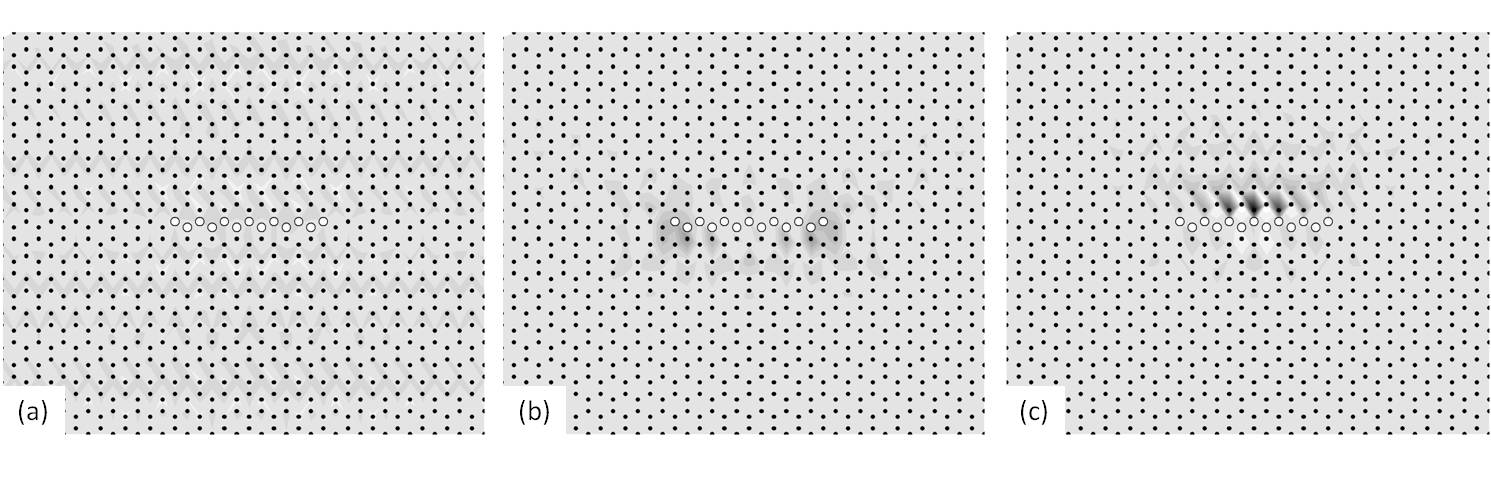}}
 \caption{Magnetic patterns in a graphene
sheet induced by a zigzag line defect placed at the center of the sheet. The
positions of the impurity atoms are denoted by white symbols. Here, we
consider various impurity scattering strengths (a) $U_{d}/t=1$, (b)$U_{d}/t=3
$, and (c) $U_{d}/t=5$, while keeping U/t=1.2. The color scale ranges from
black (negative magnetization) to white (positive magnetization). }
 \end{figure}

So far, we have focused on the limit of very strong impurity scattering,
corresponding to vacancy defects. When these vacancy sites are replaced by
impurity atoms, the resulting scattering strengths are typically smaller,
i.e. of the order $U_{d}\sim t$. In Fig. 5, we study the
evolution of the magnetic patterns induced by a zigzag line defect as a
function of increasing impurity scattering strength, while leaving U/t=1.2.

Examining the spatial structure of these magnetic patterns, we observe that
they become more localized with increasing impurity scattering strength,
which is expected. More interesting, however, is the evolution of these
patterns. For weak impurity scattering ($U_{d}/t$)=1, the local magnetic
moment is not strongly localized around the impurity sites. When the
scattering strength is increased to $U_{d}/t=3$, the local magnetic moment
localizes more strongly around the edge of the impurity cluster with maximum
induced magnetic moment are $M_{B}=-0.0187\mu _{B}$ and $M_{A}=0.0024\mu _{B}
$. For even stronger scattering, i.e. $U_{d}/t=5,$ the local magnetic moment
localizes close to the center of impurity cluster, with maximum induced
magnetic moment are $M_{B}=-0.0586\mu _{B}$ and $M_{A}=0.0088\mu _{B}$. A
Gaussian shape along the direction parallel to the zigzag defect is
observed, which was already seen in Fig.4(b) for
the case when $U_{d}/t$ tends to infinity.  When the scattering strength of
impurity increased, a stronger localized state is formed around the impurity
sites, and hence a stronger local magnetic moment.

\section{conclusion}

In summary, we have studied the  effects of point, zigzag and armchair line defects
on the electronic and magnetic structure in graphene sheets, using a self-consistent
numerical solution of the mea- field Hubbard model on a two-dimensional
honeycomb lattice. In the vicinity
of point and zigzag defects, we observe pronounced impurity induced scattering
resonances in the electronic density of states, which are largely absent for
armchair line defects. In the case of a point
vacancy defect, the amplitude of the impurity induced local density of
states is found to decay inversely proportional to $r^{2}$ , and its
frequency is observed to converge as $\left\vert E_{r}\right\vert \sim 1/U_{d}$
with increasing impurity scattering strength. The local electronic density
of states around zigzag line defects is found to be strongly enhanced as well. The
amplitudes of the impurity induced scattering resonances decay with a power
law similar to the case of a point defect, and otherwise their spatial
dependence is rather featureless, with the exception of a local minimum in
the local electronic density of states appearing near the center of zigzag
line defects, indicating destructive interference.
For linear clusters of  impurities, we observe that the induced local magnetic 
moments are enhanced close to the center of the line defect, indicating the 
formation of spin polarized localized states. Furthermore, for the case
of randomly placed defects, we find that maximum conductivity can be 
achieved at a non-trivial critical defect concentration.  
Strong impurity-induced magnetic patterns are also observed in the
vicinity of point defects and zigzag line defects. For the case of
point defects, a threefold symmetric magnetic pattern is observed. In the
case of zigzag line defects, the amplitudes of the defect induced magnetic
moments are strongest at the center of the line defect and weaker at its
ends. Generally, the impurity induced magnetic patterns of the zigzag line
defect display a Gaussian spatial pattern along the direction of the line.
This strong orientational magnetic pattern is found to persist down to
fairly small impurity scattering strengths of $U_{d}/t=5$, below which the
induced patterns become more uniform.

\begin{acknowledgments}
We would like to thank B. Normand, Tim Wehling, and Z. Y. Ming for useful discussions. Computing facilities were generously provided by the University of Southern California High-Performance Supercomputing Center. Furthermore, we acknowledge financial support by the Department of Energy under grant DE-FG02-05ER46240.
\end{acknowledgments}


\begin{thebibliography}{5}
\bibitem{note} Our fitting value W is smaller than the value reported by
Wehling (6.06eV), because the impurity configuration is slightly different.
In Ref \cite{TOW07}, the impurity only acts on one C-atom. In our case, the
impurity is embedded in the graphene sheet which forms chemical bonds with
all three nearest-neighbor carbon atoms.
\bibitem{note3} The slight top-bottom asymmetries observed at the boundaries
of Figs. 5(a)-(c) are due to finite size effects. In order to
apply periodic boundary conditions to the zigzag edges of the fonote
lattice, we had to choose an even number of lines along the zigzag
direction. In this case, the line defect cannot be perfectly centered, thus
giving rise to the slightly non-symmetric magnetic pattern.
\end{thebibliography}

\end{document}